\documentclass[%
 reprint,
 amsmath,amssymb,
 aps,
pra,
]{revtex4-2}
\usepackage{titlesec}
\titleformat{\section}{\normalfont\normalsize\bfseries}{\thesection.}{1em}{}
\usepackage{setspace} 
\newcommand{\X}{{\mathcal X}}
\usepackage{graphicx}
\usepackage{xcolor}
\usepackage{dcolumn}
\usepackage{bm}
\usepackage{comment}

\definecolor{dgreen}{rgb}{0.0, 0.66, 0.42}

\usepackage{bbm}

\begin{document}

\preprint{APS/123-QED}

\title{Maximum entropy inference of reaction-diffusion models}

\author{Olga Movilla Miangolarra$^\dagger$}
\author{Asmaa Eldesoukey$^\dagger$}
\author{Ander Movilla Miangolarra$^\ddagger$}
\author{Tryphon T. Georgiou$^\dagger$}
\affiliation{ $\dagger$
Department of Mechanical and Aerospace Engineering, University of California, Irvine, US}
\affiliation{ $\ddagger$
Department of Computational and Systems Biology, John Innes Centre, UK
}

\date{\today}

\begin{abstract}
Reaction-diffusion equations are commonly used to model a diverse array of complex systems, including biological, chemical, and physical processes. Typically, these models are phenomenological, requiring the fitting of parameters to experimental data. In the present work, we introduce a novel formalism to construct reaction-diffusion models that is grounded in the principle of maximum entropy. This new formalism aims to incorporate various types of experimental data, including ensemble currents, distributions at different points in time, or moments of such. To this end, we expand the framework of Schr\"odinger bridges and Maximum Caliber problems to nonlinear interacting systems. We illustrate the usefulness of the proposed approach by modeling the evolution of (i) a morphogen across the fin of a zebrafish and (ii) the population of two varieties of toads in Poland, so as to match the experimental data.
\end{abstract}

\maketitle



\section*{Introduction}
\vspace{-12pt}

Reaction-diffusion equations serve as dynamical models widely applicable across science. They arise naturally in chemistry but also in biology, ecology, physics, and geology. The rich behavior emerging from their nonlinear interactive nature, together with their relative simplicity, makes them a useful playing field to test theories and make predictions.
Specifically, reaction-diffusion models have been used to study biological pattern formation~\cite{kondo2010reaction},  spatial ecology \cite{holmes1994partial,cantrell2004spatial}, epidemic dynamics \cite{peng2012reaction}, tumor growth \cite{gatenby1996reaction}, 
chemotaxis \cite{stevens2000derivation}, and intracellular transport \cite{soh2010reaction}, to name a few. Not limited to modeling natural interactive systems, reaction-diffusion processes have also proven useful in recent developments in machine learning and image denoising \cite{guo2011reaction,chen2015learning,chen2016trainable}.

Due to the inherent level of complexity of the underlying physical systems,
typical reaction-diffusion models are phenomenological. In turn, the relevant parameters often need to be fitted or inferred from experimental data. Recent works have introduced multiple ways of doing so, ranging from machine learning techniques \cite{schnoerr2016cox,rao2023encoding,zhang2024discovering}, to inverse problems~\cite{soubeyrand2014parameter,matas2024unraveling}, or diverse variational approaches~\cite{hogea2008image,dewar2010parameter,srivastava2024pattern}.
Here, we take an alternative route and infer reaction-diffusion dynamics from a fundamental principle, Jaynes' {\em maximum entropy principle}~\cite{jaynes1980minimum}. 

The maximum entropy principle seeks to represent the current state of knowledge about a system via a probability distribution that is maximally non-committal to unavailable information, that is, the distribution with the largest entropy~\cite{jaynes1980minimum}.
%
%
In a dynamical context, this implies that the model that best describes the current knowledge of the dynamics of a system of interest, when no other information is available, is the one induced by the probability distribution on paths that maximizes entropy~\cite{presse2013principles}.
This idea is encapsulated in the 
 dubbed {\em Maximum Caliber principle}, which allows for incorporating various types of data, while maintaining maximum uncertainty about the remainder of the system~\cite{ghosh2020maximum,pachter2023foundations}. It has proven highly effective as an inference method, particularly in the context of complex systems~\cite{dixit2018perspective} such as~gene circuits \cite{firman2017building}, protein folding~\cite{wan2016maximum,zhou2017bridging}, or bird flocking~\cite{cavagna2014flocking}.

A related approach has been developed in parallel that dates back to E.\ Schr\"odigner's work in 1931~\cite{Sch31,Sch32}. 
Therein, Schr\"odigner showed that, given knowledge about the stochastic dynamics of a cloud of particles, the most likely evolution of the particles traversing between two observed endpoint distributions that are inconsistent with the prior dynamics, is the one that minimizes the relative entropy from the prior. It is noted that minimizing relative entropy with respect to a uniform prior is equivalent to maximizing the entropy of a distribution on paths.
Thus, this so-called {\em Schr\"odinger bridge problem}, which was originally conceived as a large deviations problem \cite{dembo2009large}, in light of Jaynes's work, can be interpreted as an inference problem. 
In addition, the Schr\"odinger bridge problem can also be seen as a stochastic optimal control problem, where particles are driven from one endpoint distribution to another through a control action of minimum energy \cite{chen2016relation}.
Whether it is seen as an inference, large deviations, or control problem, it has had a multitude of applications, from machine learning \cite{de2021diffusion} to biology~\cite{schiebinger2019optimal}, meteorology~\cite{fisher2009data}, or robotics \cite{elamvazhuthi2019mean}. 

It turns out that Maximum Caliber and Schr\"odinger bridge problems can be solved in a unified form~\cite{movilla2024inferring}. 
In the present work, we develop and apply maximum entropy principles (and in particular, Maximum Caliber and Schr\"odinger bridge theories) to nonlinear reaction-diffusion systems.
To this end, we introduce a stochastic process in which particles interact, whose ensemble behavior follows any given reaction-diffusion equation. This 
allows us to quantify distance between reaction-diffusion models via a relative entropy functional between the corresponding stochastic processes.
Leveraging tools originating from classical Schr\"odinger bridges, we obtain structured analytical solutions that are computationally approachable, as illustrated through two examples. In the first example, we adapt the dynamics of a bone morphogenetic responsive element in the pectoral fin of a zebrafish to match the gradient obtained in \cite{mateus2020bmp}. In the second, we model the evolution of the population of two varieties of toads across southern Poland to match experimental data \cite{szymura1986genetic}.

\section*{Mathematical model}
\vspace{-12pt}

Consider the reaction-diffusion equation
\begin{equation}\label{eq:RD}
    \partial_t c=\frac12\sigma^2\Delta c-\mathcal D(c)c+\mathcal A(c).
\end{equation}
Paradigmatically, $c(t,x)$ represents the density of a chemical species at time $t\in[0,T]$ and position $x\in\mathbb R$. This chemical species diffuses in a medium with diffusion coefficient $\sigma(t,x)$, while transforming back and forth into other (untracked) species with rates  $\mathcal D(t,x,c(t,x))$, $\mathcal A(t,x,c(t,x))$ that respectively model the disappearance and appearance of particles of the species of interest. Note that $\mathcal D$ and $\mathcal A$ can depend on $c(t,x)$, corresponding to a system of interacting particles. The terms in~\eqref{eq:RD} need to be bounded so that the total concentration is finite at all times, i.e., $\int c(t,x) dx<\infty$.

To develop a maximum entropy formalism for this kind of system, we have to assign a microscopic description that provides a measure of ``likelihood" to individual particle trajectories. Multiple microscopic processes may lead to the same desired ensemble behavior of the particles following \eqref{eq:RD}, so the choice should be motivated by the physics of the particular system at hand. One of the many possible choices could be a reaction-diffusion master equation, which considers the joint probability of the discrete number of particles in each discretization of space \cite{smith2019spatial}. Here, however, we consider an alternative choice that is closer in spirit to classical Schr\"odinger bridge theory: a particle-based approach where we follow the stochastic dynamics of a single particle interacting with the rest of the particles through their marginal distributions.

Specifically, since probability mass should be conserved, we consider a reservoir where the particles of the chemical species being tracked disappear to and/or appear from. 
 To this end, we let $\Omega=\mathcal S([0,T],\,\mathcal X\cup \{r\})$ be the space of right-continuous functions with left limits ({\em {c\`adl\`ag}}),  from $[0,T]$ to $\mathcal X\cup \{r\}$. Here, $\X=\mathbb R$ denotes the primary space where the chemical species of interest lives, while $\{r\}$ denotes a one-point reservoir which stores the chemical species that we are not keeping track of.
Thus, $\Omega$ represents the space of the admissible microscopic trajectories that the individual particles might take over the time window of interest. 

For a each particle, we consider a stochastic process  $\{\mathbf X_t\}$ on $\X\cup\{r\}$ that specifies a distribution $Q$ on the sample paths in $\Omega$, and denote by $Q_t$ the one-time density on $\X\cup\{r\}$. By this we mean that $Q_t(x)dx$ is the probability of $\mathbf X_t\in[x,x+dx]$, with $x\in\X$, while $Q_t(r)$ is the probability of $\mathbf{X}_t=r$. 
For specificity, we write
$$
\mathbf X_t=\begin{cases}
  X_t
  \ \mbox{ if }\mathbf X_t\in\X,
  \\R_t \  \mbox{ if }\mathbf X_t\in\{r\}.
\end{cases}
$$
That is, $\{X_t\}$
is the restriction the stochastic process $\{\mathbf X_t\}$ to $\X$, and $\{R_t\}$ the restriction of $\{\mathbf X_t\}$ to $\{r\}$. 
These obey the Markovian dynamics
\begin{subequations}
\begin{align}\label{eq:X}
    dX_t&=\sigma dB_t,\  \mbox{ and }\ 
   dR_t=0,
\end{align}
where $\{B_t\}$ denotes a standard Brownian motion. That is, when in the primary space, the particle diffuses with diffusion coefficient $\sigma$, and when in the reservoir, its dynamics are trivial.
 The process transitions between the two spaces $\mathcal X$ and $\{r\}$ with mean-field rates 
 \begin{equation}\label{eq:rates}
     \mathcal D(t,X_t,q(t,X_t)) \ \mbox{ and } \ 
\mathcal A(t,X_t,q(t,X_t)),
 \end{equation}
\end{subequations}
where $q(t,\cdot)$ denotes the 
restriction of the one-time density $Q_t$ to $\X$, i.e., $q(t,x)=Q_t(x)$ for $x\in\X$.

\begin{figure}
    \centering
\includegraphics[width=\linewidth,trim={1.95cm 1.2cm 1.425cm 1cm},clip]{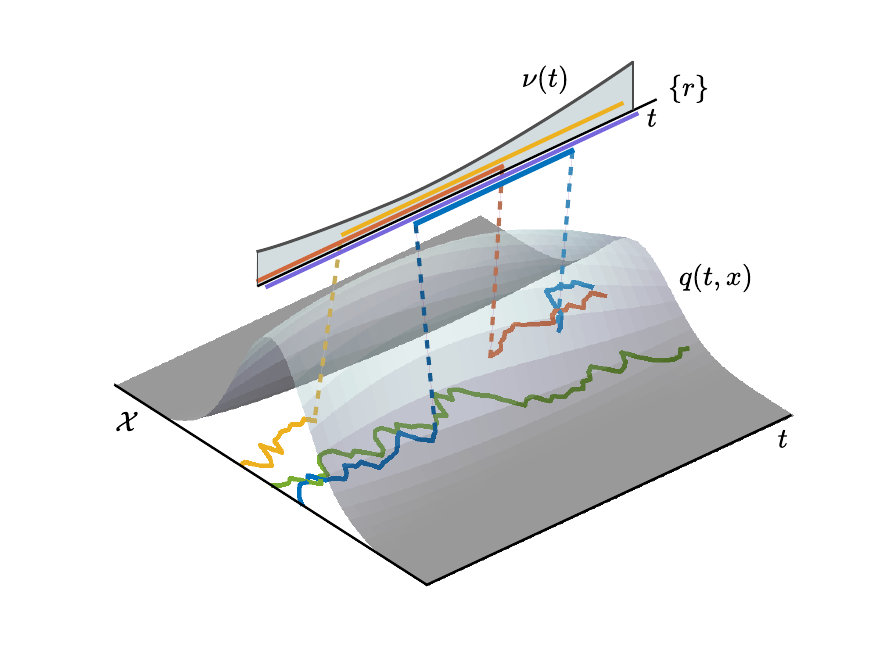}
    \caption{Schematic of sample trajectories on $\X\cup \{r\}$. Trajectories may stay for the whole duration of time at the primary space (green), at the reservoir (purple), cross over from the primary space to the reservoir (yellow), or vice-versa (red), and transition several times (blue). The one-time densities $q$ and $\nu$ are shaded in gray.}
    \label{fig:schematic}
\end{figure}

These mean-field transition rates model the rate at which particles from the primary space $\X$ migrate to the reservoir $\{r\}$, and vice-versa. That is, $\mathcal D(t,x,q(t,x))$ denotes the rate at which a particle at time $t$
 and position $x\in\X$ transitions to $\{r\}$ (i.e., transforms into a chemical species that we are not keeping track of), while $\mathcal A(t,x,q(t,x))$ denotes the rate at which a particle at $\{r\}$ at time $t$ transitions to $x\in\X$ and thus transforms into the chemical species of interest. These are mean-field rates since they are assumed to depend on the other particles of the ensemble only through the density $q(t,x).$
Therefore, the stochastic process $\{\mathbf X_t\}$ is a hybrid process that alternates between diffusive evolution and trivial evolution through a jump process (right continuous Markov process). See Figure \ref{fig:schematic} for an illustration of some example trajectories.

 \pagebreak
Letting $\nu(t)$ denote the restriction of $Q_t$ to $\{r\}$, then
 $q(t,x)$ and $\nu(t)$ satisfy
\begin{align*}
    \partial_t q&=\frac12\Delta(\sigma^2 q)-\mathcal D(q)q+\mathcal A(q)\nu,
    \\
     \partial_t \nu&=-\int_\X \mathcal A(q)\nu dx+\int_\X \mathcal D(q)q dx,
\end{align*}
where $\nu$ accounts for the mass/species we are not keeping track of. If the reservoir is much larger than our system of interest, i.e. if $\int_\X q(t,x) dx\ll \nu(t)$ at all times~$t$, then $\nu$ is approximately constant. Consequently, we may disregard the dynamical equation for $\nu$ to recover the sought ensemble behavior \eqref{eq:RD} for $q$, after a suitable rescaling $\mathcal A(q) \nu\to \mathcal A(q)$.

\vspace{-12pt}
\section*{Maximum entropy inference}

\vspace{-12pt}

Motivated by the work in \cite{backhoff2020mean}, where the authors develop a Schr\"odinger bridge theory for particles driven by mean-field potentials, we consider an analogous entropic cost to measure distances between models. Specifically, we consider
    \begin{align}
\label{eq:H(PQ)}
 H(P|Q(P))=\int_\Omega dP\log\frac{dP}{dQ(P)},
\end{align}
where $Q(P)$ denotes the distribution on paths induced by the stochastic process $\{\mathbf X_t^P\}$, with its restrictions to $\X$ and $\{r\}$ satisfying
$$
dX^P_t=\sigma dB_t \ \mbox{ and  } \ dR^P_t=0,
    $$
and transition rates between them
   $$ \mathcal D(t,X^P_t,p(t,X^P_t))
 \mbox{ and } \mathcal A(t,X^P_t,p(t,X^P_t)),
$$
respectively.
That is, $Q(P)$ is the distribution on paths of the prior process \emph{with the mean-field transition rates evaluated at the posterior distribution of particles} $p(t,x)$, where $p(t,x)$ is the restriction of $P_t$ to $\X$.
The expression~\eqref{eq:H(PQ)} is well-defined when $P$ gives zero measure to sets that have zero $Q(P)$-measure, this is, when $P$ is absolutely continuous with respect to $Q(P)$; a property that from now on is denoted as $P\ll Q(P)$.

Note that \eqref{eq:H(PQ)} is not a standard relative entropy, since $Q(P)$ depends on $P$.
The authors in~\cite{backhoff2020mean} showed that, 
when the interaction of diffusive particles is mediated through a mean-field potential drift (instead of through mean-field transition rates, as in this work), the process satisfies a large deviation principle with \eqref{eq:H(PQ)} as the rate function.
In the present work, we do not attempt to establish a large deviations theory, but rather utilize \eqref{eq:H(PQ)} as an inference/control cost.

Thus, we consider the inference problem in which we seek a distribution on paths $P$ (i.e. a dynamical model) that best describes our current knowledge. That is, one that minimizes relative entropy with respect to a prior model $Q$ in the sense of \eqref{eq:H(PQ)}, while satisfying certain empirical constraints. Specifically, we consider
    \begin{subequations}\label{eq:SBP}
    \begin{align}
&\min_{P\ll Q(P)} H(P|Q(P))\\ \label{eq:SB}&\mbox{ s.t. } \  P_0(\cdot)=P^{\rm exp}_0(\cdot),\ P_T(\cdot)=P^{\rm exp}_T(\cdot),\\\label{eq:MaxCal}
&\mbox{ and/or } \int_\Omega\int_0^Tj(t,X_t)dtdP=J^{\rm exp},
\end{align}
  \end{subequations}
where \eqref{eq:SB} fixes the endpoint distributions as in the Schr\"odinger bridge, while \eqref{eq:MaxCal}
imposes a current-constraint typical of Maximum Caliber problems.

Let us first solve the optimization problem with both constraints and then explain how to dispense with either of them.
To do so, it is key to note that for any $P\ll Q(P)$, minimizing the relative entropy \eqref{eq:H(PQ)} is equivalent to minimizing
    \begin{align}\nonumber
\int_0^T\int_\X\bigg(&\frac12\frac{u^2}{\sigma^2}p+(\delta\log\delta -\delta +1)\mathcal D(p)p\\&+(\alpha\log\alpha -\alpha +1)\mathcal A(p)\mu\bigg)dxdt,\label{eq:var-ent}
    \end{align}
    where $\mu(t)$ is the restriction of $P_t$ to $\{r\}$, while $\alpha(t,x)$, $ \delta(t,x)$ and $u(t,x)$ prescribe the posterior evolution via
\begin{subequations}\label{eq:p-mu-post}
       \begin{align}
        \partial_tp&=-\nabla\cdot(up)+\frac{1}{2}\Delta(\sigma^2 p) -\delta \mathcal D(p)p+\alpha \mathcal A(p)\mu, \\
\partial_t\mu&=-\int_\X \alpha \mathcal A(p)\mu dx+\int_\X \delta \mathcal D(p) p dx.
    \end{align}
   \end{subequations} 
    The justification of this statement is provided in Appendix \ref{app:rel-ent}.
In light of equations \eqref{eq:var-ent} and \eqref{eq:p-mu-post}, problem \eqref{eq:SBP} can be re-interpreted as that of finding the optimal controls $u,\, \alpha,$ and $\delta$ that drive the system between endpoint distributions and/or achieve a certain current constraint. Note that $u$ represents an added drift while $\alpha$ and $\delta$ modify the transition rates.


Standard calculus of variations  then leads to the following  first-order necessary conditions for optimality (see Appendix \ref{app:var-opt}),
$$
u=\sigma^2\nabla\log\varphi,\quad  \alpha=\frac{\varphi}{\psi},\quad \delta=\frac{\psi}{\varphi},
$$
with the one-time densities decomposing into
$$
p(t,x)=\varphi(t,x)\hat\varphi(t,x) \ \mbox{ and }\ \mu(t)=\psi(t)\hat\psi(t),
$$
and where $\varphi,\, \hat\varphi,\,\psi$ and $\hat\psi$ solve the following system
\begin{subequations}\label{eq:SSS}
\begin{align}\nonumber
\partial_t\varphi&=-\frac12\sigma^2\Delta\varphi+\lambda j\varphi +(\varphi-\psi )\mathcal D\\
\label{eq:varphi}&\quad\ +\varphi(\varphi-\psi )\Big(\hat\varphi\partial_p\mathcal D-\hat\psi\partial_p\mathcal A\Big),
\\\nonumber
\partial_t\hat\varphi&=\frac12\Delta(\sigma^2\hat\varphi)-\lambda j\hat\varphi-\hat\varphi\mathcal D +\hat\psi\mathcal A\\\label{eq:hatvarphi}&
\quad\ -\hat\varphi(\varphi-\psi)\big(\hat\varphi \partial_p\mathcal D-\hat\psi\partial_p\mathcal A\big) ,\end{align}
\begin{align}\label{eq:psi}
\partial_t\psi&= \int_\X (\psi-\varphi) \mathcal A dx,\\\label{eq:hatpsi}
\partial_t\hat\psi&=-\int_\X\hat\psi\mathcal Adx
 +\int_\X\hat\varphi\mathcal Ddx,
\end{align}
with endpoint conditions
\begin{align}
\label{eq:SS0}&\varphi(0,x)\hat\varphi(0,x)=p^{\rm exp}_0(x),\\   &\psi(0)\hat\psi(0)=M-\int_\X p^{\rm exp}_0(x)dx,
 \\
\label{eq:SST}&\varphi(T,x)\hat\varphi(T,x)=p^{\rm exp}_T(x),
 \\& \psi(T)\hat\psi(T)=M-\int_\X p^{\rm exp}_T(x)dx.\label{eq:SST2}
\end{align} 
\end{subequations}
Here, $M=\int_\Omega dP$ is the total amount of (probability) mass fixed by the prior, and $p^{\rm exp}_i(x)=P^{\rm exp}_i(x)$ for $x\in\X$ and $i\in\{0,T\}$, are the restrictions to the primary space of the given marginal distributions.
The scalar $\lambda$ in (\eqref{eq:varphi}-\eqref{eq:hatvarphi}) is the Lagrange multiplier corresponding to constraint \eqref{eq:MaxCal}, which needs to be suitably chosen to ensure the constraint is satisfied (see \cite[Section III. A. 3]{movilla2024inferring} for details on this for the convex case). Therefore, if we are only interested in the Schr\"odinger bridge problem, to match the two marginals, we may drop constraint \eqref{eq:MaxCal} by setting $\lambda=0.$

It is worth noting that the structure of the minimum entropy updates on the drift and transition rates $u,\,\alpha$ and $\delta$ all depend on $\varphi$. If either of them is updated, then all of them must be updated. In fact, $\alpha$ and $\delta$ are inverse of each other, i.e., $\alpha=1/\delta$ \footnote{From the point of view of discrete Schrödinger 
bridges this is not surprising. Indeed, this can be understood as a form of diagonal scaling update on the transition rates (see \cite[Eq.\ 21]{georgiou2015positive}).}.
Furthermore, the temporal symmetry characteristic of Schr\"odinger bridges is lost due to the asymmetry induced by $\mathcal D$ and $\mathcal A$. 
However, if $\mathcal D=\mathcal A$, the temporal symmetry is restored; the bridge from $P_T$ to $P_0$ is the same as the bridge from $P_0$ to $P_T$ traced backward in time.

In classical Schr\"odinger bridges, this is, when the dynamics are linear and particles do not interact ($\mathcal A,\mathcal D,\mu,\lambda=0$), the system \eqref{eq:SSS} turns into a system of linear PDEs coupled only at the boundaries, known as Schr\"odinger system. In that case, the system has a unique solution~\cite{fortet1940resolution,essid2019traversing}, which can be obtained by a convergent algorithm due to Robert Fortet \cite{fortet1940resolution}, \cite[Section~8]{chen2021stochastic}. The algorithm consists in alternating between solving \eqref{eq:hatvarphi} forward in time and \eqref{eq:varphi} backward in time, using $p_0^{\rm exp}$ and $p_T^{\rm exp}$ to obtain the initial condition for one after computing the terminal condition for the other. Schematically, this can be expressed as iterating the steps in the following diagram:
\begin{align}\label{eq:sinkhorn}\nonumber
\hat\varphi&(0,x)\quad\xrightarrow{\eqref{eq:hatvarphi}}\quad\hat\varphi(T,x)\\\tfrac{p_0^{\rm exp}(x)}{\varphi(0,x)}\ & \uparrow\qquad\qquad\qquad\qquad\downarrow\ \tfrac{p_T^{\rm exp}(x)}{\hat\varphi(T,x)}\\
\varphi&(0,x)\quad\xleftarrow{\eqref{eq:varphi}}\quad\varphi(T,x).\nonumber
\end{align}

In our case ($\mathcal A,\mathcal D,\lambda$ possibly nonzero), uniqueness of solutions to the system \eqref{eq:SSS} is not guaranteed since the optimization problem \eqref{eq:SBP} may no longer be convex due to the nonlinear nature of the system \footnote{Given \eqref{eq:H(PQ)}, convexity (and thus uniqueness) is ensured if $\mathcal D(p)$ and $\mathcal A(p)$ are such that 
$
(\alpha\log\alpha-\alpha)\mathcal A(p)\mu$ and $ 
(\delta\log\delta-\delta)\mathcal D(p)p
$
are jointly convex on $\alpha,\,p,\,\mu$ and $\delta,\,p$, respectively.
For example, $\mathcal A(p)={\rm constant}$ and $\mathcal D(p)=1/p$ are possible candidates. 
}. Operationally, one can still try to use a Fortet-type algorithm, in which we alternate between solving (\ref{eq:hatvarphi},\ref{eq:hatpsi}) forward and (\ref{eq:varphi},\ref{eq:psi}) backward in time,
to obtain a solution to the system~\eqref{eq:SSS}. Since the equations are now coupled at all times (not only at the boundaries as in the classical Schr\"odinger system), the forward equations must use the solution to the backward equation (in the previous iteration) at all times, and vice-versa. Due to the nonlinearity of the partial differential equations, there are no guarantees for the algorithm's convergence.
Numerical evidence suggests that the algorithm converges for certain prior dynamics (e.g. the KKP Fisher equation used as the prior for Figure \ref{fig:frogs}), while it is seen to fail for other more complex cases (e.g. multiple chemical species displaying Turing patterns).


If instead of imposing constraint \eqref{eq:SB} we impose \eqref{eq:MaxCal}, this is, if we have a Maximum Caliber problem instead of a Schr\"odinger bridge, then the solution must still satisfy (\ref{eq:varphi}-\ref{eq:hatpsi}) but with different endpoint conditions. The argument detailed in Appendix \ref{app:MaxCal}, which is similar in spirit to that provided in \cite{movilla2024inferring}, shows that in the Maximum Caliber case, instead of (\ref{eq:SS0}-\ref{eq:SST}), we have
\begin{subequations}\label{eq:MaxCalBC}
\begin{align}
   & \hat\varphi(0,x)=q_0(x),\quad \hat\psi(0)=M-\int_\X q_0(x)dx,\\
 &\varphi(T,x)=K,\qquad \psi(T)=K,  
\end{align}
\end{subequations}
where $K$ is a constant that ensures that the total posterior mass is $M$.
Clearly, with these endpoint conditions, iteration of the type \eqref{eq:sinkhorn} is no longer necessary, since the endpoint conditions are now uncoupled. However, given that the forward equations depend on the solution to the backward equation and vice-versa, iterating might still be necessary.  After solving (\ref{eq:varphi}-\ref{eq:hatpsi}) with \eqref{eq:MaxCalBC} as endpoint conditions, it only remains to find the appropriate $\lambda$ to satisfy the constraint~\eqref{eq:MaxCal}.

A remark on the generality of the framework is in order. 
To simplify notation, we have restricted ourselves to the one-dimensional case with a single species in $\X=\mathbb R$. Moreover, we have considered prior dynamics with no drift/advection (see \eqref{eq:X}). However, none of these restrictions are necessary; we may slightly adapt the proofs in Appendices \ref{app:rel-ent}, \ref{app:var-opt} and \ref{app:MaxCal} to account for multiple dimensions ($\X=\mathbb R^n)$, and prior dynamics with general drift terms $b(t,x,p(t,x))$ that can capture the mean-field interaction between particles \footnote{One simply needs to minimize the same relative entropy cost \eqref{eq:H(PQ)} with $(u-b)^2$ instead of $u^2$, and change $u\to b+u$ in the dynamics of $p$.}. Since adapting the framework to multiple species is less standard and more involved, we do so explicitly in Appendix \ref{app:multiple}. Finally, it is typical of reaction-diffusion systems to evolve on bounded domains, e.g. $\X=[x_{\rm min},x_{\rm max}]$, with periodic or no-flux boundary conditions. We may also deal with these cases by adapting the conditions at the boundary of $\X$ on the dynamical equations for $\varphi$ and $\hat\varphi$ \cite{caluya2021reflected}. More on this can be found at the end of Appendix \ref{app:var-opt}. Thus, postponing these (rather technical) details to the Appendices, we next illustrate the results through some examples. 

\begin{figure}[t]
    \centering
\includegraphics[width=\linewidth,trim={0.575cm 0.5cm 5.3cm 0.4cm},clip]{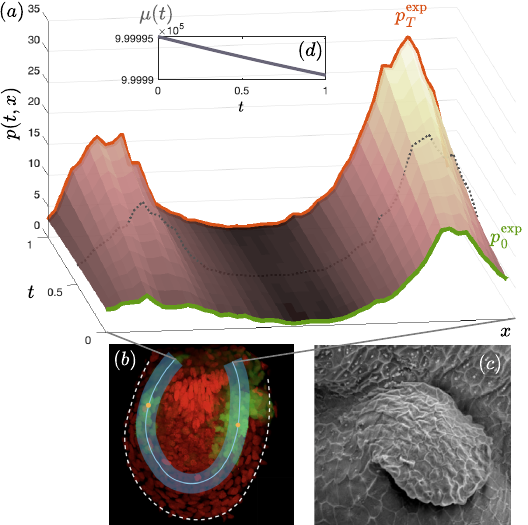}
    \caption{(a) Inferred dynamics of BRE intensity across the (straightened) region of interest of the zebrafish pectoral fin,
    given two experimental data sets at dimensionless times $t=0$ and $t=1$. The black-dotted ($t=0.5$) curve represents a third experimental data set that was not used in the inference of the model, and is displayed for comparison.
    (b)~The region of interest is shaded in light blue; a dashed white line delineates the pectoral fin. 
    (c)~The pectoral fin
    imaged through electron microscopy. 
    (d)~Mass in the (virtual) reservoir as a function of time.
    Sub-figures (b-c) have been taken from~\cite{mateus2020bmp}.}
    \label{fig:BRE}
\end{figure}

\vspace{-12pt}
\section*{Examples}
\vspace{-13pt}

Let us first consider the evolution of bone morphogenetic protein (BMP) signaling in the pectoral fin of a zebrafish as it grows. BMP signaling has been shown to regulate embryonic fin growth \cite{ning2013microrna} and is one of the most studied morphogenetic mechanisms. In \cite{mateus2020bmp}, transgenic fish carrying a fluorescently tagged BMP responsive element (BRE) were imaged, and the intensity of BRE across a region of interest of the fin was measured (the region of interest is depicted in blue in Figure~\ref{fig:BRE}~(b)). Taking their experimental data at two time points (48 and 60 hours post-fertilization), we reconstruct the dynamics of the BRE intensity (see Figure~\ref{fig:BRE} (a)). We consider linear prior dynamics (as in \cite{mateus2020bmp}) with $\mathcal D=d$, $\mathcal A=a/M$, where we have set the  dimensionless constants to $\sigma=0.01$, $d=0.1$, $a=5$, $M=10^6$, with no-flux boundary conditions. We numerically solve \eqref{eq:SSS} through iteration \eqref{eq:sinkhorn}, which in the linear case is ensured to converge \footnote{The proof of convergence for the linear case essentially follows the same argument as in \cite[Appendix B]{chen2022most}.}. It is observed in Figure \ref{fig:BRE} that our dynamical model provides a good fit of intermediate data (dotted) that was not used in our modeling, illustrating the utility of this procedure to predict intermediate measurements. In addition, our model correctly identifies the region where the protein is being synthesized, even if the prior synthesis rate is~uniform.

\begin{figure}
    \centering    \includegraphics[width=\linewidth,trim={0.25cm 0.3cm 0.925cm 0.97cm},clip]{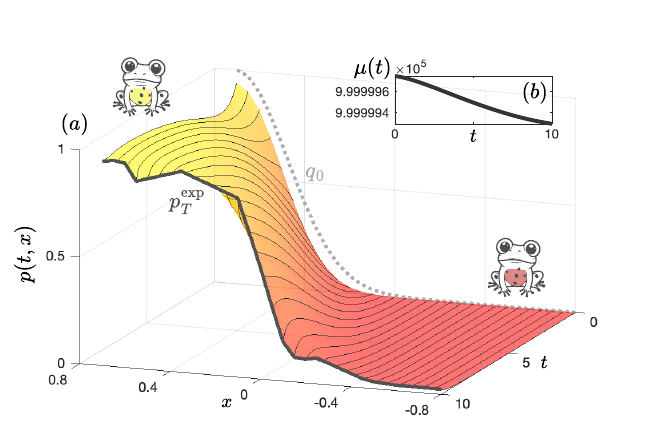}
    \caption{(a) Evolution of the average frequency of \textit{B. variegata} alleles as a function of time, with dimensionless final time T=10. The $x$-axis denotes the dimensionless distance to the hybrid zone ($x=0$), with positive $x$ denoting the \textit{B. variegata} allele dominated zone, while negative $x$ denotes the \textit{B. bombina} (fire-bellied) zone. (b) Mass at the fictitious reservoir as a function of time.}
    \label{fig:frogs}
\end{figure}

Let us consider as our second and last example the evolution of the population of two varieties of toads  (\textit{Bombina bombina} and \textit{Bombina variegata}) in southern Poland \cite{szymura1986genetic}.
A typical model for population dynamics is the KPP-Fisher equation \cite[Ch. 13]{murray2007mathematical}, i.e., equation~\eqref{eq:RD} with $\mathcal D(c)=\mathcal A(c)=rc$ for $r$ a positive constant. Taking the dimensionless constants $\sigma=0.2$ and  $r=1$ as our prior \footnote{Our prior may contain a fictitious reservoir (as opposed to the standard KKP-Fisher equation). In that case, we must normalize $\mathcal A$ to account for it. Thus, we may take $\mathcal A(c)=rc/M$ with $M$ the total mass of our system. Clearly, for large enough $M$ (e.g. $M=10^6$ as in Fig. \ref{fig:frogs}), whether a fictitious reservoir is considered or not, the resulting dynamics are the same.}, together with no-flux boundaries and the initial condition $q_0$ (displayed in Figure \ref{fig:frogs}), we infer the dynamics of the populations of these two varieties of toads to match the data obtained in \cite{szymura1986genetic}. Specifically, the available data, $p_T^{\rm exp}(x)$, is the average frequency of variegata alleles. For instance, $p(t,x)=0$ implies all alleles found at a distance $x$ from the hybrid zone $x=0$ (the zone where neither allele dominates) correspond to  \textit{B. bombina}, while $p(t,x)=1$ implies all of them are \textit{B. variegata}. 
Since we only have data at a single point in time, we simply use $q_0$ as prior data and do not impose it, seeking what is known as a half bridge \cite{pavon2021data}. 
As in the Maximum Caliber case, iteration of the form \eqref{eq:sinkhorn} is not necessary. However, we still need to solve for the coupled equations for $\varphi$ and $\hat\varphi$, one being forward in time and the other backward. To do this, we iterate between \eqref{eq:hatvarphi} with initial condition $q_0$ and \eqref{eq:varphi} with final condition $p^{\rm exp}_T(\cdot)/\hat\varphi(T,\cdot)$, obtaining the evolution displayed in Figure~\ref{fig:frogs}.
Note that in this nonlinear case, due to the lack of convexity, there are no guarantees that the numerically obtained solution constitutes a (global)~minimum.  



\section*{Conclusions}
\vspace{-12pt}

We have presented an approach to infer reaction-diffusion dynamics that accounts for empirical data in the form of ensemble averages and endpoint distributions. 
The inference relies on a notion of relative entropy for interactive particle systems \eqref{eq:H(PQ)} that is general and can be applied to models that satisfy constraints that differ from the ones considered herein.
For instance, the same optimization criterion is of value in parametric models with a known structure. In such cases, the form \eqref{eq:SSS} of solutions is lost, but the numerical search for suitable parameters is still amenable.
In other words, the framework presented here is flexible in terms of the experimental data and prior knowledge of the system, and can be easily adapted to obtain a formulation equivalent to~\eqref{eq:SSS}.

Furthermore, it would be of interest to consider alternative underlying stochastic processes realizing reaction-diffusion equations, that would lead to different notions of likelihood, such as the stochastic model in Fock spaces developed in \cite{del2022probabilistic}.
In addition, expanding this framework to formulations that allow for phase separation would make it more widely applicable to biological problems, given the increasing interest in these models during the last decade \cite{zwicker2014centrosomes,miangolarra2021steric}.

Developing algorithms to solve the Schr\"odinger (or MaxCal) system~\eqref{eq:SSS} of general nonlinear PDEs is  also of great importance, particularly for chaotic problems with multiple interacting species, for which the algorithms presented herein are seen to fail.
Generalized proximal gradient descent algorithms offer a promising route since they have been shown to converge for the mean-field Schr\"odinger bridge problem in which particles interact through the drift \cite{chen2023density}.

Lastly, one last open question concerns the theory of large deviations. In this work, we have proposed a method to infer reaction-diffusion models based on a maximum entropy principle. We have also shown that this amounts to a control problem in which particles are steered to meet a target (see \eqref{eq:H(PQ)}).
We expect the employed cost functional to be the rate function characterizing large deviations for this system of interacting particles, but the theoretical underpinnings of such a statement remain open.


\noindent\emph{Acknowledgements:} Supported in part by NSF under ECCS-2347357, AFOSR under FA9550-24-1-0278, and ARO under W911NF-22-1-0292.FA9550-23-1-0096. \\
\emph{Contact:} omovilla@uci.edu, aeldesouky@uci.edu, movilla@nbi.ac.uk, tryphon@uci.edu.\\
\emph{Data availability:} The code used herein is available at Github (https://github.com/olga-m-m/reac-diffSBs). Previously published data were
used for this work~\cite{mateus2020bmp,szymura1986genetic}.



\bibliography{arXiv}

\begin{widetext}
\newpage

\appendix

\noindent 
\textbf{ \large Appendix}\\[-0.3in]
\section{Relative entropy characterization (proof of \eqref{eq:var-ent})}
\label{app:rel-ent}
Let
$\{Y_t\}$ denote the canonical process $Y_t=\omega_t$, with $\omega\in\Omega$. 
Consider a partition $\{ \Delta_i \mid i\in\mathbb Z\}$
of $\mathcal X=\mathbb R$ into disjoint sets of vanishingly small size $dx$, and define $\mathcal P:=\{ \Delta_i \mid i\in\mathbb Z\}\cup \{r\}$. Given a collection $S_0,\ldots, S_{N}\in \mathcal P$, define the cylinder set of paths
$$
   {\mathcal S}_{S_0,\ldots, S_{N}} := \{\omega\in\Omega :  Y_0(\omega)\in S_0,Y_\epsilon(\omega)\in S_{\epsilon}, \ldots, Y_{N \epsilon}(\omega)\in S_{N}\},
$$
where
$N\epsilon=T$. Denote by $\mathcal C_\epsilon$ the collection of all such cylinder sets that coarse-grain the time axis into $N$ segments.
For $t>s$, let $\pi_{y|x}(s,t)$  denote the probability (under $P$) of $\mathbf X_t=y$ given that $\mathbf X_s=x$, i.e. the transition kernel of $P$. 
Then, the probability of the family of trajectories ${\mathcal S}_{S_0,\ldots, S_{N}}$ under $P$ can be expressed as
\begin{align*}
 dP({\mathcal S}_{S_0,\ldots, S_{N}}) = P_0(x_0) \pi_{x_{\epsilon}\mid x_{0}} (0,\epsilon) \times  \cdots \times  \pi_{x_{N\epsilon}\mid x_{(N-1) \epsilon}} ((N-1) \epsilon,N\epsilon)dx_0\ldots dx_{N\epsilon} ,
\end{align*}
where $x_{i\epsilon}\in S_i$ and $|S_i|=dx_{i\epsilon}$ for $i\in\{0,\ldots,N\}$. Note that in a slight abuse of notation, when $x_{i\epsilon}=r$, i.e., the reservoir state, then $dx_{i\epsilon}:=1.$ 

Expressing the probability mass of cylinder sets similarly for $Q(P)$, we write the discretized (in time) relative entropy between $P$ and $Q(P)$ in terms of these path probabilities as follows:
\begin{align*}
H_\epsilon(  P |  Q(P)) &= \int_{\mathcal S\in \mathcal C_\epsilon} dP( {\mathcal S}) \log \frac{ dP( {\mathcal S})}{ d Q( {\mathcal S})} \\
    & = \int_{(\X\cup\{r\})^{N+1}} P_0(x_0) \pi_{x_{\epsilon}\mid x_{0}} (0,\epsilon) \times  \cdots \times  \pi_{x_{N\epsilon}\mid x_{(N-1) \epsilon}} ((N-1) \epsilon,N\epsilon) \\
    &\quad \times \bigg \{ \log \frac{P_0(x_0)}{Q_0(x_0)} + \log \frac{ \pi_{x_{\epsilon}\mid x_{0}} (0,\epsilon)}{\rho_{x_{\epsilon}\mid x_{0}} (0,\epsilon)} + \cdots + \log \frac{\pi_{x_{N\epsilon}\mid x_{(N-1) \epsilon}} ((N-1) \epsilon,N\epsilon)}{\rho_{x_{N\epsilon}\mid x_{(N-1) \epsilon}} ((N-1) \epsilon,N\epsilon)} \bigg \} dx_0\ldots dx_{N\epsilon} ,
\end{align*}
where $\rho$ denotes the transition kernel of $Q(P).$
Since the term $\int P_0(x_0) \log \frac{P_0(x_0)}{Q_0(x_0)}dx_0$ is fixed by the initial condition~\eqref{eq:SB}, 
minimizing the relative entropy is equivalent to minimizing 
\begin{align} \label{eq:KLd} \nonumber
    &\int_{(\X\cup\{r\})^{N+1}} P_0(x_0) \pi_{x_{\epsilon}\mid x_{0}} (0,\epsilon) \times  \cdots \times  \pi_{x_{N\epsilon}\mid x_{(N-1) \epsilon}} ((N-1) \epsilon,N\epsilon) \bigg \{ \sum_{k=0}^{N-1} \log \frac{\pi_{x_{(k+1)\epsilon}\mid x_{k \epsilon}} (k \epsilon,(k+1)\epsilon)}{\rho_{x_{(k+1)\epsilon}\mid x_{k \epsilon}} (k \epsilon,(k+1)\epsilon)} \bigg \} dx_0\ldots dx_{N\epsilon}
     \\
   &= \sum_{k=0}^{N-1}\int_{(\X\cup\{r\})^2} P_{k\epsilon}(x_{k\epsilon}) \pi_{x_{(k+1)\epsilon} \mid x_{k\epsilon}} (k \epsilon,(k+1)\epsilon)  \log \frac{    \pi_{x_{(k+1)\epsilon} \mid x_{k\epsilon}} (k \epsilon,(k+1)\epsilon) }{  \rho_{x_{(k+1)\epsilon}\mid x_{k \epsilon}} (k \epsilon,(k+1)\epsilon)} dx_{k\epsilon} dx_{(k+1)\epsilon}.
\end{align}

For the above expression to be bounded, it is necessary that
$\pi_{x_{(k+1)\epsilon}\mid x_{k \epsilon}} (k \epsilon,(k+1)\epsilon)=0$ whenever $\rho_{x_{(k+1)\epsilon}\mid x_{k \epsilon}} (k \epsilon,(k+1)\epsilon)=0$. This amounts to considering only absolutely continuous distributions $P$ with respect to $Q(P)$.
Thus, $P$ must only differ from $Q(P)$ by a drift $u(t,x)$ (when in the primary space) and positive multiplicative functions $\alpha(t,x), \, \delta(t,x)$ on the discrete transition rates (when transforming from one species to another).
Specifically, each transition can be of one of four types:\\
(i) If $x_{k\epsilon},\, x_{(k+1)\epsilon}\in\X$, 
\begin{align*}
     \pi_{x_{(k+1)\epsilon}\mid x_{k \epsilon}} (k \epsilon,(k+1) \epsilon)&=\frac{1}{\sqrt{2\pi\epsilon\sigma_{k\epsilon}^2}}e^{-(x_{(k+1)\epsilon}-x_{k \epsilon}-u_{k \epsilon}\epsilon)^2/(2\epsilon\sigma_{k\epsilon}^2)}\big(1-\epsilon \delta_{(k+1)\epsilon}\mathcal D_p\big((k+1)\epsilon,x_{(k+1)\epsilon}\big)+o(\epsilon)\big),
\end{align*}
where we have used the condensed notations $\delta_{(k+1)\epsilon}:=\delta((k+1)\epsilon,x_{(k+1)\epsilon})$, $u_{k\epsilon}:=u(k\epsilon,x_{k\epsilon})$, $\sigma_{k\epsilon}:=\sigma(k\epsilon,x_{k\epsilon})$, and $\mathcal D_p\big((k+1)\epsilon,x_{(k+1)\epsilon}\big):=\mathcal D\big((k+1)\epsilon,x_{(k+1)\epsilon},p((k+1)\epsilon,x_{(k+1)\epsilon})\big)$.\\
(ii) If $x_{k\epsilon}\in\X$ and $x_{(k+1)\epsilon}=r$, then
\begin{align}\nonumber
     \pi_{r\mid x_{k \epsilon}} (k \epsilon,(k+1) \epsilon)&=\int_\X\frac{1}{\sqrt{2\pi\epsilon\sigma_{k\epsilon}^2}}e^{-(x-x_{k \epsilon}-u_{k \epsilon}\epsilon)^2/(2\epsilon\sigma_{k\epsilon}^2)}\epsilon\delta((k+1)\epsilon,x)\mathcal D_p((k+1)\epsilon,x)dx+o(\epsilon)
     \\\nonumber &=\epsilon\delta((k+1)\epsilon,x_{k\epsilon})\mathcal D_p\big((k+1)\epsilon,x_{k\epsilon})\big)+o(\epsilon).
\end{align}
(iii) If $x_{k\epsilon}=r$ and $x_{(k+1)\epsilon}\in\X$,
\begin{align*}
     \pi_{x_{(k+1)\epsilon}\mid r} (k \epsilon,(k+1) \epsilon)&=\epsilon\alpha_{(k+1)\epsilon}\mathcal A_p\big((k+1)\epsilon,x_{(k+1)\epsilon}\big) +o(\epsilon),
\end{align*}
where  $\alpha_{(k+1)\epsilon}:=\alpha((k+1)\epsilon,x_{(k+1)\epsilon})$ and $\mathcal A_p\big((k+1)\epsilon,x_{(k+1)\epsilon}\big) :=\mathcal A\big((k+1)\epsilon,x_{(k+1)\epsilon},p((k+1)\epsilon,x_{(k+1)\epsilon})\big)$.
\\
(iv) If $x_{k\epsilon}= x_{(k+1)\epsilon}=r$ 
\begin{align*}
     \pi_{r\mid r} (k \epsilon,(k+1) \epsilon)&=1-\epsilon\int_\X\alpha((k+1) \epsilon,x)\mathcal A_p\big((k+1) \epsilon,x\big) dx +o(\epsilon).
\end{align*}
Therefore, 
(i) if $x_{k\epsilon},\, x_{(k+1)\epsilon}\in\X$, 
\begin{align*}
    \log\frac{    \pi_{x_{(k+1)\epsilon} \mid x_{k\epsilon}} (k \epsilon,(k+1)\epsilon) }{  \rho_{x_{(k+1)\epsilon}\mid x_{k \epsilon}} (k \epsilon,(k+1)\epsilon)}=&\log\left(\frac{e^{-(x_{(k+1)\epsilon}-x_{k \epsilon}-u_{k \epsilon}\epsilon)^2/(2\epsilon\sigma_{k\epsilon}^2)}}{e^{-(x_{(k+1)\epsilon}-x_{k \epsilon})^2/(2\epsilon\sigma_{k\epsilon}^2)}}\right)+\log\left(\frac{1-\epsilon \delta_{(k+1)\epsilon}\mathcal D_p\big((k+1)\epsilon,x_{(k+1)\epsilon}\big)+o(\epsilon))}{1-\epsilon \mathcal D_p\big((k+1)\epsilon,x_{(k+1)\epsilon}\big)+o(\epsilon)}\right)
     \\=& \frac{1}{\sigma_{k\epsilon}^2}\Big(u_{k \epsilon}(x_{(k+1)\epsilon}-x_{k \epsilon})-   \frac12\epsilon u_{k \epsilon}^2\Big)+\log\big(1+\epsilon (1-\delta_{(k+1)\epsilon})\mathcal D_p\big((k+1)\epsilon,x_{(k+1)\epsilon}\big)+o(\epsilon)\big)
     \\=&
 \frac{1}{\sigma_{k\epsilon}^2}\Big(u_{k \epsilon}(x_{(k+1)\epsilon}-x_{k \epsilon})-   \frac12\epsilon u_{k \epsilon}^2\Big)
    +\epsilon(1-\delta_{(k+1)\epsilon})\mathcal D_p\big((k+1)\epsilon,x_{(k+1)\epsilon}\big)+o(\epsilon),
\end{align*}
(ii) if $x_{k\epsilon}\in\X$ and $x_{(k+1)\epsilon}=r$,
\begin{align*}
        \pi_{r\mid x_{k\epsilon}} (k \epsilon,(k+1)\epsilon) \log\frac{    \pi_{r\mid x_{k\epsilon}} (k \epsilon,(k+1)\epsilon) }{  \rho_{r\mid x_{k\epsilon}} (k \epsilon,(k+1)\epsilon)}=&\epsilon\delta\big((k+1)\epsilon,x_{k\epsilon}\big)\mathcal D_p\big((k+1)\epsilon,x_{k\epsilon}\big)\log\delta((k+1)\epsilon,x_{k\epsilon})+o(\epsilon),
\end{align*}
(iii) if $x_{k\epsilon}=r$ and $x_{(k+1)\epsilon}\in\X$,
\begin{align*}
      \pi_{x_{(k+1)\epsilon} \mid r} (k \epsilon,(k+1)\epsilon)   \log\frac{    \pi_{x_{(k+1)\epsilon} \mid r}  (k \epsilon,(k+1)\epsilon) }{  \rho_{x_{(k+1)\epsilon} \mid r}  (k \epsilon,(k+1)\epsilon)}=\epsilon\alpha_{(k+1)\epsilon}\mathcal A_p((k+1)\epsilon,x_{(k+1)\epsilon} )\log(\alpha_{(k+1)\epsilon})+o(\epsilon),
\end{align*}
and (iv) if $x_{k\epsilon}= x_{(k+1)\epsilon}=r$, 
\begin{align*}
      &\pi_{r \mid r} (k \epsilon,(k+1)\epsilon) \log\frac{    \pi_{r \mid r} (k \epsilon,(k+1)\epsilon) }{  \rho_{r\mid r} (k \epsilon,(k+1)\epsilon)}\\&=\left(1-\epsilon\int_\X\alpha((k+1) \epsilon,x)\mathcal A_p((k+1) \epsilon,x) dx +o(\epsilon)\right)\log\left(\frac{1-\epsilon\int_\X\alpha((k+1) \epsilon,x)\mathcal A_p((k+1) \epsilon,x) dx  +o(\epsilon)}{1-\epsilon\int_\X\mathcal A_p((k+1) \epsilon,x) dx +o(\epsilon)}\right)\\&
      =\epsilon\int_\X\mathcal (1-\alpha({(k+1)\epsilon},x))\mathcal A_p((k+1)\epsilon,x) dx +o(\epsilon).
\end{align*}

By splitting the summation in \eqref{eq:KLd} into the different cases (i-iv), using $k\in{\rm (i)}$ as shorthand for $\{k: x_{k\epsilon},\, x_{(k+1)\epsilon}\in\X\}$ and similarly for (ii-iv), we obtain
   \begin{align*}
   H_\epsilon(  P | Q(P))&= \sum_{k\in{\rm(i)}}\int_{\X^2} p(k\epsilon,x_{k\epsilon}) \pi_{x_{(k+1)\epsilon} \mid x_{k\epsilon}} (k \epsilon,(k+1)\epsilon)   \\
   &\hspace{1cm} \times\bigg(\frac{1}{\sigma_{k\epsilon}^2}u_{k \epsilon}(x_{(k+1)\epsilon}-x_{k \epsilon})-   \frac{1}{2\sigma_{k\epsilon}^2}\epsilon u_{k \epsilon}^2 +\epsilon(1-\delta_{(k+1)\epsilon})\mathcal D_p\big((k+1)\epsilon,x_{(k+1)\epsilon}\big)\bigg)dx_{k\epsilon} dx_{(k+1)\epsilon} 
   \\
   &+\sum_{k\in{\rm (ii)}}\int_{\X} p(k\epsilon,x_{k\epsilon}) \bigg(\epsilon\delta((k+1)\epsilon,x_{k\epsilon})\mathcal D_p\big((k+1)\epsilon,x_{k\epsilon}\big)\log\delta((k+1)\epsilon,x_{k\epsilon})\bigg) dx_{k\epsilon}  \\
   &+\sum_{k\in{\rm (iii)}} \int_{\X}\mu({k\epsilon})\bigg(\epsilon\alpha_{(k+1)\epsilon}\mathcal A_p\big((k+1)\epsilon,x_{(k+1)\epsilon} \big)\log(\alpha_{(k+1)\epsilon})\bigg) dx_{(k+1)\epsilon} 
    \\&+\sum_{k\in{\rm (iv)}} \mu({k\epsilon})\bigg(\epsilon\int_\X\mathcal (1-\alpha({(k+1)\epsilon},x))\mathcal A_p\big((k+1)\epsilon,x\big) dx \bigg)  +o(\epsilon).
\end{align*}

The first terms, as $\epsilon\to 0$, become
\begin{align*}
  &\sum_{k\in {\rm (i)}}   \int_{\X^2} p(k\epsilon,x_{k\epsilon}) \pi_{x_{(k+1)\epsilon} \mid x_{k\epsilon}} (k \epsilon,(k+1)\epsilon) \frac{1}{\sigma_{k\epsilon}^2}\left(u_{k \epsilon}(x_{(k+1)\epsilon}-x_{k \epsilon})-   \frac12\epsilon u_{k \epsilon}^2\right)dx_{k\epsilon} dx_{(k+1)\epsilon}  
  \to  \frac12 \int_0^T\int_\X \frac{u^2}{\sigma^2} p dxdt,
\end{align*}
where we used the fact that $x_{(k+1)\epsilon}-x_{k \epsilon}$ equals $u_{k \epsilon}\epsilon$ plus a martingale term whose expectation vanishes. 
Similarly, the next term of the sum becomes
$$
\sum_{k\in {\rm (i)}}\int_{\X^2} p(k\epsilon,x_{k\epsilon}) \pi_{x_{(k+1)\epsilon} \mid x_{k\epsilon}} (k \epsilon,(k+1)\epsilon)  \Big( \epsilon(1-\delta_{(k+1)\epsilon})\mathcal D_p\big((k+1)\epsilon,x_{(k+1)\epsilon}\big)\Big) dx_{k\epsilon} dx_{(k+1)\epsilon}\to \int_0^T\int_\X (1-\delta)\mathcal D(p) p dxdt.
$$
The term corresponding to (ii) turns
$$\sum_{k\in{\rm (ii)}}\int_{\X} p(k\epsilon,x_{k\epsilon}) \bigg(\epsilon\delta((k+1)\epsilon,x_{k\epsilon})\mathcal D_p((k+1)\epsilon,x_{k\epsilon})\log\delta((k+1)\epsilon,x_{k\epsilon})\bigg) dx_{k\epsilon}   \to
\int_0^T\int_\X   \delta\log(\delta) \mathcal D(p)p dx dt.
$$
Similarly, the term corresponding to (iii) becomes
$$
\sum_{k\in{\rm (iii)}} \int_{\X}\mu({k\epsilon})\bigg(\epsilon\alpha_{(k+1)\epsilon}\mathcal A_p((k+1)\epsilon,x_{(k+1)\epsilon}) \log(\alpha_{(k+1)\epsilon})\bigg) dx_{(k+1)\epsilon} \to\int_0^T\int_\X   \alpha\log(\alpha) \mathcal A(p)\mu dx dt.
$$
Finally, as $\epsilon \to 0$, the term corresponding to (iv) turns
$$
\sum_{k\in{\rm (iv)}} \mu({k\epsilon})\bigg(\epsilon\int_\X\mathcal (1-\alpha({(k+1)\epsilon},x))\mathcal A_p((k+1)\epsilon,x) dx \bigg) \to \int_0^T\int_\X(1-\alpha)\mathcal A(p)\mu dxdt.
$$
Putting all the ingredients together we obtain the desired result. This same proof can be adapted for the general case with $N$ species to yield \eqref{eq:H(P,Q(P))N} in Appendix \ref{app:multiple} below.

\section{\label{app:var-opt}Optimization problem (proof of \eqref{eq:SSS})}

We would like to solve the following problem
\begin{align}\nonumber
&\min_{u,p,\delta,\alpha}\int_0^T\int_\X\left(\frac12\frac{u^2}{\sigma^2}p+(\delta\log\delta -\delta +1)\mathcal D(p)p+(\alpha\log\alpha -\alpha +1)\mathcal A(p)\mu\right)dxdt\\\nonumber
& \mbox{ s.t. } \partial_tp=-\nabla\cdot (up)+\frac{1}{2}\Delta (\sigma^2 p) -\delta \mathcal D(p)p+\alpha \mathcal A(p)\mu, \quad   p(0,x)=p^{\rm exp}_0(x),\ p(T,x)=p^{\rm exp}_T(x),\quad \int_0^T\int_\X jpdxdt=J^{\rm exp},\\
&\qquad\ \partial_t\mu=-\int_\X \alpha \mathcal A(p)\mu dx+\int_\X \delta \mathcal D(p) p dx, \quad \mu(0)=M-\int_\X p_0^{\rm exp}dx,\ \mu(T)=M-\int_\X p^{\rm exp}_Tdx\nonumber.
\end{align}
The Lagrangian for this optimization problem reads
\begin{align*}
\mathcal L=\int_0^T\bigg[\int_\X &\bigg(
\frac12\frac{u^2}{\sigma^2}p+(\delta\log\delta -\delta +1)\mathcal D(p)p+(\alpha\log\alpha -\alpha +1)\mathcal A(p)\mu+\lambda j p\\&+\log\varphi\Big(\partial_t p+\nabla\cdot(up)-\frac{1}{2}\Delta (\sigma^2 p)-\alpha \mathcal A(p)\mu+\delta \mathcal D(p)p\Big)+\log\psi\Big(\alpha \mathcal A(p)\mu-\delta \mathcal D(p) p
\Big)\bigg)dx+\log\psi\partial_t\mu \bigg] dt,
\end{align*}
where $\lambda$, $\log\varphi(t,x)$ and $\log\psi(t)$ are Lagrange multipliers.
Rearranging the terms and integrating by parts we obtain
\begin{align}\nonumber
\mathcal L=\int_0^T&\bigg\{\int_\X\bigg(
\frac12\frac{u^2}{\sigma^2}p+\Big(\delta\log\frac{\delta\varphi}{\psi} -\delta +1\Big)\mathcal D(p)p+\Big(\alpha\log\frac{\alpha\psi}{\varphi} -\alpha +1\Big)\mathcal A(p)\mu+\lambda j p\\\nonumber
&\qquad-(\partial_t\log\varphi) p
-\frac12 \sigma^2 (\Delta\log\varphi)p
-(\nabla\log\varphi) u p\bigg)dx-(\partial_t\log\psi)\mu+BC,
\end{align}
where we have the following boundary terms from the integration by parts
\begin{align}
    \label{eq:boundary}
BC=&\int_0^T\left[\log\varphi(t,x)u(t,x)p(t,x)-\frac12\log\varphi(t,x)\nabla \big(\sigma(t,x)^2 p(t,x)\big)+\frac12\sigma(t,x)^2p(t,x)\nabla \log\varphi(t,x)\right]^{x\to\infty}_{x\to-\infty}
 dt\\&
+\left[\int_\X p(t,x)\log\varphi(t,x)dx+\mu(t)\log\psi(t)\right]^{t=T}_{t=0}.\nonumber
\end{align}
The temporal boundary terms are fixed by the initial and final densities, while the spatial boundary terms vanish assuming that $p(t,x),\nabla p(t,x)\to 0$ as $x\to \pm\infty$.

Taking the first variation of $\mathcal L$ with respect to $u(t,x)$ and setting it to zero, we obtain
$$
\frac{1}{\sigma^2}up-\nabla\log\varphi p=0\ \implies\ u=\sigma^2\nabla\log\varphi.
$$
Similarly with respect to $\alpha$  and $\delta$
\begin{align*}
   \log\frac{\alpha\psi}{\varphi}\mathcal A(p)\mu&=0\ \implies \alpha=\frac{\varphi}{\psi},\\ \log\frac{\delta\varphi}{\psi}\mathcal D(p)p&=0\ \implies \delta=\frac{\psi}{\varphi}.
\end{align*}
Using these three expressions for $u$, $\alpha$ and $\delta$, we obtain that the first variation with respect to $p(t,x)$ leads to
\begin{align*}
    &-\partial_t\log\varphi-\frac12\sigma^2(\nabla\log\varphi)^2-\frac12\sigma^2\Delta\log\varphi+\Big(1-\frac{\psi}{\varphi}\Big)(\mathcal D(p)+p\partial_p\mathcal D(p))+\Big(1-\frac{\varphi}{\psi}\Big)\partial_p\mathcal A(p)\mu+\lambda j=0\\&\implies \partial_t\varphi=-\frac12\sigma^2\Delta\varphi+\lambda j\varphi+(\varphi-\psi )\mathcal D(p)+\varphi(\varphi-\psi )(\hat\varphi\partial_p\mathcal D(p)-\hat\psi\partial_p\mathcal A(p)),
\end{align*}
where we have used the identity $\Delta\log \varphi+(\nabla \log \varphi)^2=\Delta \varphi/\varphi$, and have defined $\hat\psi=\mu/\psi$ and $\hat\varphi=p/\varphi.$
On the other hand, the first variation with respect to $\mu(t)$ gives \begin{align}\label{eq:number}
    &-\partial_t\log\psi+\int_\X\big(1-\frac{\varphi}{\psi}\big)\mathcal Adx=0\\\nonumber
    &\implies\ \partial_t\psi= \int_\X (\psi-\varphi) \mathcal Adx.
\end{align}
Taking the time derivative of $\hat\varphi=p/\varphi$ and $\hat\psi=\mu/\psi$ one can check that $\hat\varphi$ and $\hat\psi$ follow, respectively,
\begin{align*}
\partial_t\hat\varphi&=\frac12\Delta(\sigma^2\hat\varphi)-\lambda j \hat\varphi-\hat\varphi\mathcal D+\hat\psi\mathcal A-\hat\varphi(\varphi-\psi )(\hat\varphi\partial_p\mathcal D(p)-\hat\psi\partial_p\mathcal A(p)),\\
\partial_t\hat\psi&=-\int_\X\hat\psi\mathcal Adx+\int_\X\hat\varphi\mathcal Ddx,
\end{align*}
as in \eqref{eq:SSS}.

This same proof (as well as the proof in Appendix \ref{app:rel-ent}) can be adapted to more general cases. First, the proof can be adapted to the case with $N$ species to yield \eqref{eq:SSSm}. Moreover, it is also possible to account for a drift term $b(t,x)$ in the prior dynamics by changing $u^2\to(u-b)^2$ in the cost functional. This drift term might even be the mean-field result of interaction between particles, yielding a nonlinear dependence on $p$, $b(t,x,p(t,x)),$ as in \cite{backhoff2020mean}.
In addition, we can adjust this framework to consider $\X=[x_{\rm min},x_{\rm max}]\subset \mathbb R$ with the posterior density $p$ satisfying either periodic or no-flux conditions at the boundary of $\X$. 
In fact, having a bounded domain with periodic or no-flux boundary conditions is typically preferred when modeling reaction-diffusion systems (see the examples in the main text).

Periodic boundary conditions can be easily accounted for by identifying $x_{\rm min}$ with $x_{\rm max}$ and defining a Brownian motion in that space (one-dimensional circle). This leads to the same optimality equations~\eqref{eq:SSS},  with different conditions at the boundary of $\X$. Specifically, imposing periodicity of $p$ removes the boundary terms \eqref{eq:boundary}, implying that  $\varphi$ and $\hat\varphi$ can be chosen to be periodic on $\X$.
On the other hand, no-flux boundary conditions, i.e., for either $x_{\rm min}$ or $x_{\rm max}$,
\begin{equation}\label{eq:no-flux}
    J(t,x_{\rm min/max}):=-u(t,x_{\rm min/max})p(t,x_{\rm min/max})+\frac12\nabla \big(\sigma(t,x_{\rm min/max})^2p(t,x_{\rm min/max})\big)=0,
\end{equation} 
require a little more care 
\cite{caluya2021reflected}.
In this case, the optimizer follows the same first-order necessary conditions \eqref{eq:SSS}, but with $\varphi$ and $\hat\varphi$ satisfying 
$$\nabla \varphi(t,x_{\rm min/max})=0 \mbox{ and }\nabla \big(\sigma(t,x_{\rm min/max})^2\hat\varphi(t,x_{\rm min/max})\big)=0,
$$
at the boundary of $\X$.
To see this, note that condition \eqref{eq:no-flux} makes the two first boundary terms in~\eqref{eq:boundary} cancel each other, while the third term need not be zero (in this, $x$ has to be evaluated at $x_{\rm min/max}$ instead of $\pm \infty$). The first variations of the free parameters $p(t,x_{\rm min})$ and $p(t,x_{\rm max})$ lead to $\nabla\log\varphi(t,x_{\rm min/max})=0$. This implies that $u(t,x_{\rm min/max})=0$, and $\nabla \varphi(t,x_{\rm min/max})=0$. Thus, using \eqref{eq:no-flux} we obtain that $\nabla \big(\sigma(t,x_{\rm min/max})^2p(t,x_{\rm min/max})\big)=0$. Since, $\nabla (\sigma^2p)=\sigma^2\hat\varphi\nabla\varphi+\varphi\nabla(\sigma^2\hat\varphi)$, we have that $\hat\varphi$ must also satisfy the no-flux boundary condition, which in this case is simply $\nabla \big(\sigma(t,x_{\rm min/max})^2\hat\varphi(t,x_{\rm min/max})\big)=0.$ Note that the stochastic process associated with a no-flux boundary condition must be contained in the bounded domain; this is not straightforward but can be done by introducing a local time process~\cite{caluya2021reflected}.

\section{Maximum Caliber endpoint conditions (proof of \eqref{eq:MaxCalBC})}\label{app:MaxCal}

From the argument presented in Appendix \ref{app:var-opt}, it is clear that if the initial and terminal constraints -- the Schrödinger bridge constraints -- are absent, the sought distribution $P$ will still be such that its one-time densities $p$ and $\mu$ satisfy~\eqref{eq:p-mu-post} for the optimal choices
\begin{equation}\label{eq:opt-choice}
    u=\sigma^2\nabla\log\varphi,\quad \delta=\frac{\psi}{\varphi},\quad\alpha=\frac{\varphi}{\psi},
\end{equation}
 where $\varphi,\,\hat\varphi,\,\psi,\,\hat\psi$ follow (\ref{eq:varphi}-\ref{eq:hatpsi}). 
However, the endpoint conditions will differ from (\ref{eq:SS0}-\ref{eq:SST2}).

 To find the appropriate endpoint conditions, let us use the discretized paths, kernels, and notation introduced in Appendix \ref{app:rel-ent} to write the Radon-Nikodym derivative of $P$ with respect to $Q(P)$ as
\begin{align*}
    \frac{dP}{dQ(P)}({\mathcal S}_{S_0,\ldots, S_{N}})=&\frac{P_0(x_0)}{Q_0(x_0)}\frac{\pi_{x_{\epsilon}\mid x_{0}} (0,\epsilon)}{\rho_{x_{\epsilon}\mid x_{0}} (0,\epsilon)} \times  \cdots \times  \frac{\pi_{x_{N\epsilon}\mid x_{(N-1) \epsilon}} ((N-1) \epsilon,N\epsilon)}{\rho_{x_{N\epsilon}\mid x_{(N-1) \epsilon}} ((N-1) \epsilon,N\epsilon)}.
\end{align*}
As before, if (i) $x_{k\epsilon},\, x_{(k+1)\epsilon}\in\X$,
\begin{align*}
  \frac{    \pi_{x_{(k+1)\epsilon} \mid x_{k\epsilon}} (k \epsilon,(k+1)\epsilon) }{  \rho_{x_{(k+1)\epsilon}\mid x_{k \epsilon}} (k \epsilon,(k+1)\epsilon)}&
  =e^{(x_{(k+1)\epsilon}-x_{k \epsilon})u_{k \epsilon}/\sigma_{k\epsilon}^2-u^2_{k \epsilon}\epsilon/(2\sigma_{k\epsilon}^2)}\Big(1+\epsilon(1-\delta_{(k+1)\epsilon})\mathcal D_p((k+1)\epsilon,x_{(k+1)\epsilon})\Big)+o(\epsilon)  
  \\&=e^{(x_{(k+1)\epsilon}-x_{k \epsilon})u_{k \epsilon}/\sigma_{k\epsilon}^2-u^2_{k \epsilon}\epsilon/(2\sigma_{k\epsilon}^2)+\epsilon(1-\delta_{(k+1)\epsilon})\mathcal D_p((k+1)\epsilon,x_{(k+1)\epsilon})}+o(\epsilon),  
\end{align*}
 if (ii) $x_{k\epsilon}\in\X,\, x_{(k+1)\epsilon}=r$,
 \begin{align*}
  \frac{    \pi_{r \mid x_{k\epsilon}} (k \epsilon,(k+1)\epsilon) }{  \rho_{r\mid x_{k \epsilon}} (k \epsilon,(k+1)\epsilon)}&=\delta((k+1)\epsilon,x_{k\epsilon})+o(\epsilon),
  \end{align*}
 if (iii) $x_{k\epsilon}=r,\, x_{(k+1)\epsilon}\in\X$,
 \begin{align*}
  \frac{    \pi_{x_{(k+1)\epsilon} \mid r} (k \epsilon,(k+1)\epsilon) }{  \rho_{x_{(k+1)\epsilon}\mid r} (k \epsilon,(k+1)\epsilon)}&=\alpha((k+1)\epsilon,x_{(k+1)\epsilon})+o(\epsilon),
  \end{align*}
and if (iv) $x_{k\epsilon}=x_{(k+1)\epsilon}=r$,
\begin{align*}
  \frac{    \pi_{r \mid r} (k \epsilon,(k+1)\epsilon) }{  \rho_{r\mid r} (k \epsilon,(k+1)\epsilon)}&
=1+\epsilon\int_\X\Big(1-\alpha((k+1) \epsilon,x)\Big)\mathcal A_p((k+1) \epsilon,x) dx +o(\epsilon)  
  \\&=e^{\epsilon\int_\X(1-\alpha((k+1) \epsilon,x))\mathcal A_p((k+1) \epsilon,x) dx }+o(\epsilon). 
\end{align*}
Putting these together, using the fact that $x_{(k+1)\epsilon}-x_{k \epsilon}\xrightarrow[]{\epsilon\to 0} udt+\sigma dB_t $, and defining $\mathbbm{1}_S$ to be the characteristic function on $S$, we can take the limit of the Radon-Nikodym derivative as $\epsilon\to 0$ to obtain 
\begin{align}
   \frac{dP}{dQ(P)}({\mathcal S}_{S_0,\ldots, S_{N}})\xrightarrow[]{\epsilon\to 0}&\ \frac{P_0(x_0)}{Q_0(x_0)}e^{\int_0^T\mathbbm 1_{\X}(x_t)\big(\frac{1}{2\sigma^2}u^2dt+\frac1\sigma udB_t+(1-\delta)\mathcal Ddt\big)}e^{\int_0^T\mathbbm 1_{\{r\}}(x_t)\int_\X(1-\alpha)\mathcal Adxdt}\prod_{d\in \tau_d}\delta(d,x_{d})\prod_{a\in \tau_a}\alpha(a,x_{a}),  \label{eq:RN}
\end{align}
where $\tau_d$ and $\tau_a$ are the set of times at which the trajectory  transitions from $\X$ to $\{r\},$ and from  $\{r\}$ to $\X$, respectively.

The It\^o rule for $\log\varphi$ together with \eqref{eq:varphi} can be used to show that
$$
   \sigma\nabla\log\varphi dB_t=d\log\varphi -\left(\lambda j+\Big(1-\frac{\psi}{\varphi}\Big)\mathcal D+(\varphi-\psi)(\hat\varphi\partial_p\mathcal D-\hat\psi\partial_p\mathcal A)+\frac12\sigma^2(\nabla\log\varphi)^2\right)dt.
$$
Then, substituting the expression for $\sigma\nabla \log\varphi dB_t=\frac{1}{\sigma}udB_t$ in \eqref{eq:RN}, and using \eqref{eq:number} and \eqref{eq:opt-choice}, we can write 
\begin{align*}\nonumber
    \frac{dP}{dQ(P)}
=&\frac{P_0(x_0)}{Q_0(x_0)}e^{\int_0^T\mathbbm 1_{\X}(x_t)\left(d\log\varphi-\left(\lambda j+(\varphi-\psi)(\hat\varphi\partial_p\mathcal D-\hat\psi\partial_p\mathcal A)\right)dt\right)}e^{\int_0^T\mathbbm 1_{\{r\}}(x_t)d\log\psi }\prod_{d\in \tau_d}\frac{\psi(d)}{\varphi(d,x_d)}\prod_{a\in \tau_a}\frac{\varphi(a,x_{a})}{\psi(a)}.
   \end{align*}
   The telescopic expansion of $e^{\int_0^T\mathbbm 1_{\X}(x_t)d\log\varphi}$
   and 
   $e^{\int_0^T\mathbbm 1_{\{r\}}(x_t)d\log\psi}$ cancels the last two products,
  to give
   \begin{align}
    \frac{dP}{dQ(P)}
= &\frac{P_0(x_0)}{Q_0(x_0)}e^{-\int_0^T\mathbbm 1_{\X}(x_t)\left[\lambda j+(\varphi-\psi)(\hat\varphi\partial_p\mathcal D-\hat\psi\partial_p\mathcal A)\right]dt}\frac{\Phi(T,x_T)}{\Phi(0,x_0)}\nonumber\\=&\frac{\hat\Phi(0,x_0)}{Q_0(x_0)}e^{-\int_0^T\mathbbm 1_{\X}(x_t)\left[\lambda j+(\varphi-\psi)(\hat\varphi\partial_p\mathcal D-\hat\psi\partial_p\mathcal A)\right]dt}\Phi(T,x_T),\nonumber
\end{align}
where we have defined $\Phi$ ($\hat\Phi$) such that $\Phi(t,x)=\varphi(t,x)$ ($\hat\Phi(t,x)=\hat\varphi(t,x)$) if $x\in\X$, and $\Phi(t,x)=\psi(t)$ ($\hat\Phi(t,x)=\hat\psi(t)$) if $x\in\{r\}$, so that $P_t(x)=\Phi(t,x)\hat\Phi(t,x)$.

Therefore, to solve the MaxCal problem
$$
\min_{P\ll Q(P)}H(P|Q(P))\quad \mbox{s.t. } \int_\Omega\int_0^Tj(t,x)dtdP=J^{\rm exp},
$$
 it only remains to fix the optimal $\hat\Phi(0,x_0)$ and $\Phi(T,x_T)$. From a distribution of paths perspective,
since this problem does not contain any constraint on the marginals at initial and final times, it is clear that the optimal $\frac{dP}{dQ(P)}$ will not be an explicit function of $x_0$ or $x_T$. Thus,  we conclude that
$$
\hat\Phi(0,x)=Q_0(x)\quad\mbox{and}\quad\Phi(T,x)=K,
$$
where $K$ is simply a normalizing factor that ensures the total posterior mass is equal to the total prior mass $M$.




\section{Maximum entropy principle for multiple species}\label{app:multiple}
In general, we may have $N$ different species that may interact with each other, transforming from one to another at some rate. We consider, in analogy to the main text, a stochastic process $\{\mathbf X_t\}$ on $\Omega=\mathcal S([0,T],\mathbb X)$,
where we have defined $\mathbb X=\cup_{i\in\mathcal I}\X_i\cup\{r\}$, with $\mathcal I:=\{1,\ldots,N\}$ and $\X_i=\mathbb R$ for all $i$. Let 
$$
\mathbf X_t=\begin{cases}
  X_t^{(i)}
  \ \mbox{ if }\mathbf X_t\in\X_i,
  \\R_t \  \mbox{ if }\mathbf X_t\in\{r\},
\end{cases}
$$
follow, for $i\in\mathcal I$,
$$
dX_t^{(i)}=\sigma_i dB^{(i)}  \mbox{ and } dR_t=0.
$$
 The transition between the $N+1$ spaces (spaces of the $N$ species plus the common reservoir) is mediated through density-dependent transition rates. Specifically,
 $$
 \mathcal K_{i,j}(t,x,q^{(1)}(t,x),\ldots,q^{(N)}(t,x)) \mbox{ for any }i,j\in\mathcal I_0:=\{0,1,\ldots,N\}
 $$
 denotes the transition rate (at time $t$ and position $x$) from species $j$ to species $i$, with $i=0$ ($j=0$) denoting the untracked reservoir species. 
 Here, $q^{(i)}$ denotes the one-time density of $X_t^{(i)}$, with $q^{(0)}\equiv\nu$, the density at the reservoir.
These one-time densities satisfy
the following set of reaction-diffusion equations: for all $i\in\mathcal I$,
\begin{align*}
    \partial_t q^{(i)}&=\frac12\Delta(\sigma_i q^{(i)}) 
+\sum_{j\in\mathcal I_{0\backslash i}}\mathcal{K}_{i,j}\big(q^{(1)},\ldots,q^{(N)}\big)q^{(j)}-\mathcal K_{j,i}\big(q^{(1)},\ldots,q^{(N)}\big)q^{(i)} ,
\end{align*}
 where  $\mathcal I_{0\backslash i}:=\{0,1,\ldots,i-1,i+1,\ldots,N\}$, and
$$
   \partial_t\nu=\sum_{i\in\mathcal I}\int_{\X_i} \Big(\mathcal K_{0,i}\big(q^{(1)},\ldots,q^{(N)}\big) q^{(i)}-\mathcal K_{i,0}\big(q^{(1)},\ldots,q^{(N)}\big)\nu\Big) dx_i.$$
  If at all times $\nu(t) \gg \int_{\X_i} q^{(i)}(t,x_i) dx_i$ for all $i\in\mathcal I$, we can disregard the $\nu$ dynamical equation by taking $q^{(0)}=1$ and suitably rescaling $\mathcal K_{i,0}$.

Let $Q$ denote the distribution on paths of $\{\mathbf X_t\}$, and $P$ be an absolutely continuous distribution with respect to $Q(P)$, which is defined in analogy to the single species case. Denote by $P_t$ the one-time density of $P$, with  $P_t(x)=p^{(i)}(t,x)$ if $x\in\X_i$, and $P_t(x)=\mu(t)$ if $x=r$. Then, we consider the corresponding relative entropy,
\begin{align}
&H(P,Q(P))=\int_{\mathbb X} \log\frac{P_0}{Q_0}P_0dx+\sum_{i\in\mathcal I}\int_0^T\int_{\X_i}\frac12\frac{u_i^2}{\sigma_i^2}p^{(i)} dx_idt+
\sum_{i\neq j\in\mathcal I_0}\int_0^T\int_{\X_i}(\kappa_{ij}\log\kappa_{ij} -\kappa_{ij}+1)\mathcal K_{i,j}p^{(j)}dx_idt,\label{eq:H(P,Q(P))N}
\end{align}
where $p^{(0)}\equiv\mu$, and $\kappa_{ij}(t,x),$ $u_i(t,x)$
define the posterior evolution as
\begin{subequations}\label{eq:dyn-i}
    \begin{align}\partial_tp^{(i)}&=-\nabla\cdot (u_ip^{(i)})+\frac{1}{2}\Delta(\sigma_i^2 p^{(i)} )+\sum_{j\in\mathcal I_{0\backslash i}}\kappa_{ij}\mathcal{K}_{i,j}p^{(j)}-\kappa_{ji}\mathcal K_{j,i}p^{(i)} ,\\
    \partial_t\mu&=\sum_{i\in\mathcal I}\int_{\X_i} \Big(\kappa_{0i}\mathcal K_{0,i} p^{(i)}-\kappa_{i0}\mathcal K_{i,0}\mu\Big) dx_i.
\end{align}
\end{subequations}

Thus, we would like to optimize:
\begin{align}\nonumber
&\min_{u_i,p^{(i)},\kappa_{ij},\mu} H(P,Q(P))
\\\nonumber&
\mbox{ s.t. }\eqref{eq:dyn-i},\ P_0(\cdot)=P^{\rm exp}_0(\cdot),\ P_T(\cdot)=P^{\rm exp}_T(\cdot),\\\label{eq:MaxCalm}
&\mbox{ and/or } \sum_{i\in\mathcal I}\int_0^T\int_{\X_i}j_i(t,x_i)p^{(i)}(t,x_i)dx_idt=J^{\rm exp}.
\end{align}
Following the same steps as before (i.e. building a Lagrangian, taking its first derivative, and setting it to zero), we obtain the following first-order necessary conditions:
\begin{align*}
& u_i=\sigma^2_i\nabla\log\varphi_i, \ \mbox{ for all }i\in\mathcal I,\\&
      \kappa_{ij}=\frac{\varphi_i}{\varphi_j},\ \mbox{ for all }i\neq j\in\mathcal I_0, \mbox{ where }\varphi_0\equiv\psi,\ \hat\varphi_0\equiv\hat\psi,
\end{align*}
with the posterior distribution satisfying
$$
p^{(i)}(t,x)=\varphi_i(t,x)\hat\varphi_i(t,x) \mbox{ and }\mu(t)=\psi(t)\hat\psi(t).
$$
Here, $\varphi_i,\,\hat\varphi_i,\psi$ and $\hat\psi$ follow, for all $i\in\mathcal I$, 
\begin{subequations}\label{eq:SSSm}
    \begin{align}
 \partial_t\varphi_i=&-\frac12\sigma^2_i\Delta\varphi_i+\lambda j_i\varphi_i+\sum_{j\in\mathcal I_{0\backslash i}}(\varphi_i-\varphi_j)\mathcal K_{j,i} +\sum_{j\neq k\in\mathcal I_0}p^{(j)}\Big(1-\frac{\varphi_k}{\varphi_j}\Big)\varphi_i
\partial_{p^{(i)}}\mathcal K_{k,j},\\
\partial_t\hat\varphi_i=&\frac12\Delta(\sigma^2_i\hat\varphi_i)-\lambda j_i\hat\varphi_i+\sum_{j\in\mathcal I_{0\backslash i}}(\mathcal K_{i,j}\hat\varphi_j-\mathcal K_{j,i}\hat\varphi_i)
-\sum_{j\neq k\in\mathcal I_0}p^{(j)}\Big(1-\frac{\varphi_k}{\varphi_j}\Big)
\hat\varphi_i \partial_{p^{(i)}}\mathcal K_{k,j},\\
 \partial_t\psi=&\sum_{i\in\mathcal I}\int_{\X_i}
(\psi-\varphi_i)
\mathcal K_{i,0} dx_i,\\\partial_t\hat\psi=&\sum_{{i\in\mathcal I}}\int_{\X_i}(\hat\varphi_i\mathcal K_{0,i}-\hat\psi \mathcal K_{i,0})dx_i,
\end{align}
with endpoint conditions
\begin{align}
  \label{eq:SSphi0-i}&\varphi_i(0,\cdot)\hat\varphi_i(0,\cdot)=p^{{\rm exp},i}_0(\cdot),
 \\
\label{eq:SSphiT-i}&\varphi_i(T,\cdot)\hat\varphi_i(T,\cdot)=p^{{\rm exp},i}_T(\cdot),
 \\
\label{eq:SSpsi0-i}&\psi(0)\hat\psi(0)=M-\sum_{i\in\mathcal I}\int_{\X_i} p^{{\rm exp},i}_0(x_i)dx_i,
 \\
\label{eq:SSpsiT-i}&\psi(T)\hat\psi(T)=M-\sum_{i\in\mathcal I}\int_{\X_i} p^{{\rm exp},i}_T(x_i)dx_i, 
\end{align}
\end{subequations}
where $\lambda$ is the Lagrange multiplier associated with \eqref{eq:MaxCalm}.

\end{widetext}

\end{document}